# Publication Trend in DESIDOC Journal of Library and Information Technology during 2013-2017: A Scientometric Approach

Dr. M. Sadik Batcha[1], S. Roselin Jahina[2], and Muneer Ahmad[3]
[1](Associate Professor, Department of Library and Information Science, Annamalai University, Tamil Nadu, India)
[2,3](Research Scholar, Department of Library and Information Science, Annamalai University, Tamil Nadu, India)

***Abstract:*** *DESIDOC Journal of Library & Information Technology (DJLIT) formerly known as DESIDOC Bulletin of Information Technology is a peer-reviewed, open access, bimonthly journal. This paper presents a Scientometric analysis of the DESIDOC Journal. The paper analyses the pattern of growth of the research output published in the journal, pattern of authorship, author productivity, and, subjects covered to the papers over the period (2013-2017). It is found that 227 papers were published during the period of study (2001-2012). The maximum numbers of articles were collaborative in nature. The subject concentration of the journal noted is Scientometrics. The maximum numbers of articles (65 %) have ranged their thought contents between 6 and 10 pages. The study applied standard formula and statistical tools to bring out the factual result.*

## I. INTRODUCTION

DESIDOC Journal of Library & Information Technology (DJLIT) formerly (DESIDOC Bulletin of Library and Information Technology Journal) is an important journal in the field of library and information science (LIS) and is being published since 1981. It is a peer-reviewed, bimonthly, open access journal that brings to the notice of readers the recent developments in information technology (IT), as applicable to library and information science. It is meant for librarians, documentation and information professionals, researchers, students and others interested in the field. The Journal covers original research and review articles related to IT which is applied to library activities, services and products. The Journal is indexed by Scopus, LISA, LISTA, EBSCO Abstracts/Full-text, Library Literature and Information Science Index/Full-text, The Informed Librarian Online, DOAJ, Open J-Gate, Indian Science Abstracts, Indian Citation Index, Full text Sources Online, World Cat, Proquest, and OCLC.

The term scientometrics was invented by the Russian mathematician Vasiliy Nalimov (naukometriya in Russian, meaning the study of the evolution of science through the measurement of scientific information) (Nalimov and Mulchenko, 1969)[1] (Godin B)[2] This term was not noticed in Western scientific circles until it was translated into English (Garfield, 2009)[3]. Scientometrics is a science about science (Price, 1961, 1963)[4]. It provides the researchers with various concepts, models, and techniques that may be applied to any discipline in order to explore its foundations, state, intellectual core, and potential future development.(Sadik Batcha M)[5]. Many studies have analyzed the scientific and technological disciplines from a scientometric perspective, (Godin B)[6] but a brief overview of these works reveals that research output on publications lights on the trends, strength and weakness of any discipline (Moravcsik M J and Ziman S M 1975)[7] (Gunasekaran et al.2006)[8]. The Scietometrics study measures the performance based on several parameters, country annual growth rate and collaborative index. (Baskaran C and Sadik Batcha)[9]. The present investigation conducts a meta analysis of DESIDOC research Journal in order to consolidate Scientometric research on this burning field, to develop recommendations for future scientometric researchers and to better understand the identity of this scientific field. This study aims to find out the growth pattern, core journals, authorship pattern and productive authors in this field. The scientometric study of Amsaveni[10,] and Sadik Batcha[11] corroborate with the citation analysis and they imply with the impact on recent researches.

## II. REVIEW OF LITERATURE

Alka Bansal (2013)[12] in her paper presents a bibliometric analysis of the journal to assess the pattern of growth of the research output published in the journal, pattern of authorship and geographic distribution of output, subjects covered and citation analysis of the references attached to the papers and change in them over two different periods (2001- 2006) and (2007-2012). It is found that 391 papers were published during the period of study (2001-2012). The maximum number of articles (65) was published in 2012. The maximum number of contributions is joint collaborations with 61.4 %. Most of the contributions (88 %) are from India and 12 % are foreign contributions. The study revealed that majority of the authors preferred journals as the source of information providing the highest number of citations.

Bala, Madhu and Singh, Mahender Pratap (2014)[13] critically analysed 316 scholarly communications published in the Indian Journal of Biochemistry & Bio-Physics. Indian Journal of Bio-Chemistry and Bio-Physics, formerly known as IJBB. It is a peer reviewed, open access bio-monthly Journal published by NISCAIR. The analysis covered mainly the number of articles, form of document cited, most cited Journals etc. Study reveals that single author contributed 18 (5.7%) while the rest of 162 (51.3%) articles were contributed by Multi authors. The contributions in this Journal from India are slightly more than those from the other countries. The objectives of this study were to assist the collection development in order to fulfil the needs of scientists and research scholars in the field of science and technology.

P. Rajendran, R.Jeyshankar and B.Elango (2011)[14] did Scientometric analysis of 633 research articles published in Journal of Scientific and Industrial Research has been carried out. Five Volumes of the journal containing 60 issues from 2005 – 2009 have been taken into consideration for the present study. The number of contributions, authorship pattern & author productivity, average citations, average length of articles, average keywords and collaborative papers has been analyzed. Out of 633 contributions, only 51 are single authored and rest by multi authored with degree of collaboration 0.92 and week collaboration among the authors. Pattern of Co-Authorship revealed that the improving trend of co-authored papers. The study revealed that the author productivity is 0.34 and dominated by the Indian authors.

Sangeeta Paliwal (2015)[15] in her paper analysed 177 research papers published in five volumes 56 to 60, (2009 - 2013) in Annals of library and Information studies. The study gives status of Library and information science research & importance of library science in India. Also gives account of Annals of library and Information studies, objectives & methodology in this study. Analyses papers into year wise distribution, length of articles, use of tables, graphs diagrams. Finds authorship pattern and calculates collaboration coefficients. Also finds out profile contributors, location of papers, subject wise distribution and State wise distribution.

**Objectives**

The following objectives have been framed for the present study:
To identify the number of contributions and the pattern of growth of articles published in the DESIDOC journal during the period from 2013 to 2017.
- To study the Authorship Pattern exists in the published articles.
- To determine the Degree of Collaboration of the articles published during the period of study.
- To analyse the author productivity and their Exponential Growth Rate
- To identify the Subject wise distribution of contributions,
- To determine the year wise output and extent of its coverage in terms of length of articles

### III. METHODOLOGY

The present study is an effort to make it an update by studying the volumes from 33 to 37 (2013-2017). The papers and references given in these 5 issues of DESIDOC Vol. 33 - 37 (2013 - 2017) have been taken for analysis in this paper. The analysis includes 227 research articles. A data sheet was created on different aspects for main articles. The data were collected from the website of DESIDOC Journal of Library & Information Technology (DJLIT). The following statistical tools were used in the present study.
1. Relative Growth Rate and Doubling Time
2. Degree of Collaboration
3. Collaborative Index
4. Compound Annual Growth Rate
5. Exponential Growth Rate

### IV. ANALYSIS AND DISCUSSION

The collected data of 227 research articles covered in five volumes from 33 to 37 have been analysed using the statistical tools given in the methodology. There have been six issues in every volume found and totally 30 issues have been taken for analysis.

**Analysis of Year wise distribution of Articles Published**
**Analysis of the study**

**Table1: Year wise distribution of Number of Articles Published**

| S.No | Year | Volume No | Issues | No.Of.Papers | % | Cum.No.Of.Papers | Cum. % |
|---|---|---|---|---|---|---|---|
| 1 | 2013 | 33 | 6 | 33 | 14.5 | - | - |
| 2 | 2014 | 34 | 6 | 63 | 27.7 | 96 | 42.29 |
| 3 | 2015 | 35 | 6 | 44 | 19.4 | 140 | 61.67 |
| 4 | 2016 | 36 | 6 | 36 | 15.9 | 176 | 77.53 |
| 5 | 2017 | 37 | 6 | 51 | 22.5 | 227 | 100 |
|  | TOTAL | 5 | 30 | 227 | 100 |  |  |

**Graph1: Distribution of Number of Articles Published**

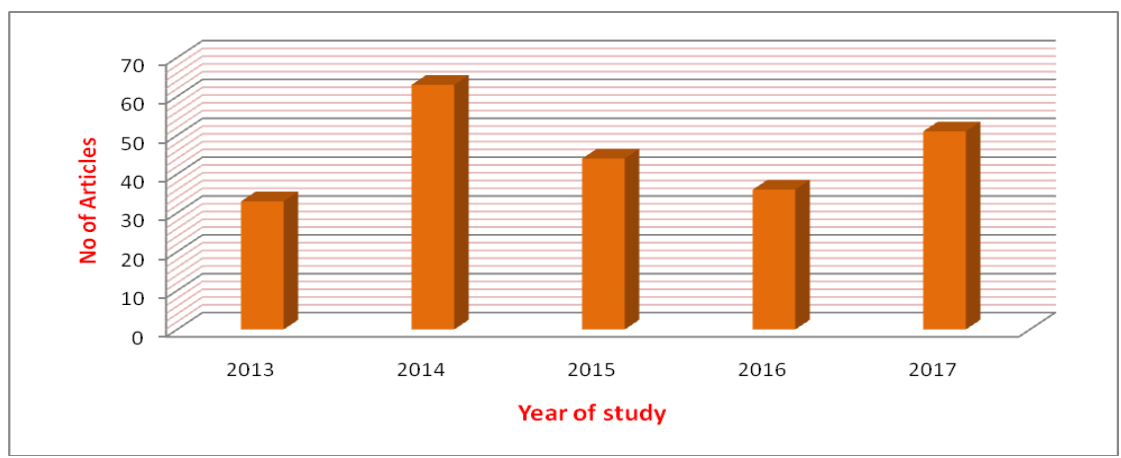

The table and Graph explain the Number of distribution of papers according to year wise. It is shown that a total of 227 papers have been published during 2013-2017. In which Maximum number of paper were published in the year 2014 which is accounted to. 63 ( 27.7%), whereas the minimum count of 33 papers were published in the year 2013 it is calculated about 14.5 percentages.

**Analysis of Authorship Pattern**

**Table2: Year wise Authorship Pattern**

| S. No | Year/Vol | Single Author | Two Author | Three Author | Four Author | Five & Above | Total | % |
|---|---|---|---|---|---|---|---|---|
| 1 | 2013/33 | 14 | 14 | 5 | - | - | 33 | 14.5 |
|  |  | 42.42 | 42.42 | 15.15 | 0.00 | 0.00 |  |  |
| 2 | 2014/34 | 21 | 28 | 9 | 5 | - | 63 | 27.8 |
|  |  | 33.33 | 44.44 | 14.29 | 7.94 | 0.00 |  |  |
| 3 | 2015/35 | 11 | 22 | 9 | 1 | 1 | 44 | 19.4 |
|  |  | 25.00 | 50.00 | 20.45 | 2.27 | 2.27 |  |  |
| 4 | 2016/36 | 12 | 17 | 5 | 1 | 1 | 36 | 15.9 |
|  |  | 33.33 | 47.22 | 13.89 | 2.78 | 2.78 |  |  |
| 5 | 2017/37 | 12 | 30 | 6 | 1 | 1 | 51 | 22.5 |
|  |  | 23.53 | 58.82 | 11.76 | 1.96 | 1.96 |  |  |
| TOTAL |  | 70 | 111 | 34 | 9 | 3 | 227 | 100 |
|  |  | 30.84 | 48.90 | 14.98 | 3.96 | 1.32 |  |  |

The table 2 explains the authorship pattern found in the published volumes of DESIDOC journals during the period of study. The study has revealed that researchers are sharing their experience producing more in Joint authorship in particular two authors are found maximum in numbers which is accounted to 111 and in percentages it is about 48.90. Yet the sole authorship is equivalently appreciable. It is noted that about 70 articles are brought out by single authorship and it is at the second of order. The three author contributions are at the third place in order followed by four and five. Five and above authors have shown only 3 articles.

**Analysis of Author Productivity**

**Table3 : Year Wise Author Productivity**

| Year | Total No. of Papers | Total No. of Authors | AAPP | Productivity Per Author |
|---|---|---|---|---|
| 2013 | 33 | 57 | 1.73 | 0.58 |
|  | 14.54 | 12.93 |  |  |
| 2014 | 63 | 124 | 1.97 | 0.51 |
|  | 27.75 | 28.12 |  |  |
| 2015 | 44 | 91 | 2.07 | 0.48 |
|  | 19.38 | 20.63 |  |  |
| 2016 | 36 | 70 | 1.94 | 0.51 |
|  | 15.86 | 15.87 |  |  |
| 2017 | 51 | 99 | 1.94 | 0.51 |
|  | 22.47 | 22.45 |  |  |
| TOTAL | 227 | 441 | 9.65 | 2.59 |

*AAPR – Average Author Per Paper

Table.3 shows the data related to author's productivity during the period of study. The total number of papers doubled from 33 to 63 in the years 2013 to 2014. It gradually decreased in the next two years i.e 2015 and 2016. Yet 2017 witnessed a growth of 51 in numbers. On the other hand the number of authors increased on par with the number of articles. Even though the number of articles calculated 44 in 2015, the average author per paper is highly shown 2.07 in this year. The total average number of authors per paper observed is 9.65 and the average productivity per author calculated is 2.59. The highest number of author's productivity found in the study was 124 (0.51%) in the year 2014. The minimum number of author's productivity noted was 57 (0.58%) in the year 2013.

**Analysis of Degree of Collaboration**

**Table4: Year Wise Degree of Collaboration**

| Year | Single Author | Multiple Author | Total Aouthors | CI | DC |
|---|---|---|---|---|---|
| 2013 | 14 | 19 | 33 | 1.36 | 0.58 |
| 2014 | 21 | 42 | 63 | 2.00 | 0.67 |
| 2015 | 11 | 33 | 44 | 3.00 | 0.75 |
| 2016 | 12 | 24 | 36 | 2.00 | 0.67 |
| 2017 | 12 | 38 | 50 | 3.17 | 0.76 |
| TOTAL | 70 | 157 | 227 | 2.24 | 0.69 |

*CI- Collaborative Index , **DC- Degree of Collaboration

**Graph2: Degree of Collaboration**

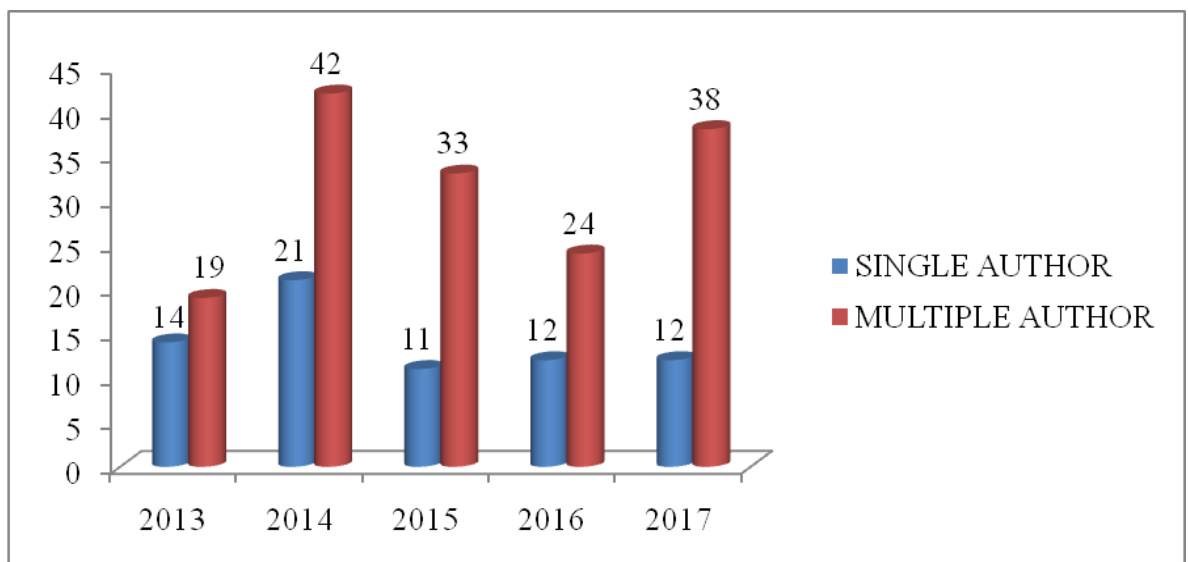

Table 4 and figure 4.1 shows the degree of author collaboration analysed in the study undertaken. It was calculated using Subramanian's formula: C = NM / (NM + NS) Where C = degree of collaboration, Nm=Number of multi-authored works, Ns= Number of single-authored works. The Collaborative Index is also calculated and presented.

$$\text{Collaborative Index} = CI = \frac{\text{No of Authors}}{\text{No. of Papers}}$$

It is found that the degree of author collaboration in the DESIDOC Journal of Library and Information Technology ranged from 0.58 to 0.76 during the period under study. In comparison, Where C = degree of collaboration 227 (0.69%), Nm=Number of multi-authored works 157, Ns= Number of single-authored works 70. The collaborative index witnessed the growth in collaboration ranged from 1.36 to 3.17.

**Analysis of Exponential Growth Rate**

**Table5: Year Wise Exponential Growth Rate**

| S.NO | YEAR | PUBLICATION | EXPONENTIAL GROWTH RATE | CAGR |
|------|------|-------------|-------------------------|------|
| 1 | 2013 | 33 | 0.00 | 9.1% |
| 2 | 2014 | 63 | 1.91 | |
| 3 | 2015 | 44 | 0.70 | |
| 4 | 2016 | 36 | 0.82 | |
| 5 | 2017 | 51 | 1.42 | |
|   | TOTAL | 227 | 4.85 | |

* CAGR – Compound Annual Growth Rate

Table- 5 shows that the Exponential Growth Rate of publications published in DESIDOC Journal of Library and Information Technology during the period of 2013 to 2017 (5 years). The highest growth rate 1.91 was found during 2014 with 63 Publications and it decreased to 0.70 and 0.82 in the next two years yet there is a steep growth noted in the next current year 2017 at the rate of 1.42. It is also found that the total Exponential Growth Rate was found to be 4.85 and CAGR calculated for five years is to be 9.1%.

**Analysis of Relative Growth of Rate and Doubling Time**

**Table6: Year wise Relative Growth of Rate and Doubling Time**

| Year | No. of publication | Cum. No. of Publication | W1 | W2 | R (a) (W1-W2) | Mean R (a) 1-2 | Doubling Time | M Dt (a)1-2 |
|------|-------------------|------------------------|------|------|---------------|----------------|---------------|-------------|
| 2013 | 33 | - | 3.49 | 4.14 | 0.65 | | 1.07 | |
| 2014 | 63 | 96 | 4.14 | 3.78 | 0.36 | | 1.93 | |
| 2015 | 44 | 140 | 3.78 | 3.58 | 0.2 | 0.61 | 3.47 | 1.78 |
| 2016 | 36 | 176 | 3.58 | 3.93 | 0.35 | | 1.98 | |
| 2017 | 51 | 227 | 3.93 | 5.42 | 1.49 | | 0.47 | |
| Total | 227 | | | | | | | |

Table – 6 indicates that the relative growth rates of articles output and also its doubling time of the publication. It could be observed that the relative growth rates of all sources of research output have decreased from 0.65 in 2013 to 0.36 in 2014. Then it revealed the peak growth rate of 1.49 in the year 2017. The mean relative growth rates for the periods 2013-2017 is observed to 0.61. The study period has witnessed a mean relative growth rate at an appreciable level. The doubling time for publication has decreased from 1.07 in 2013 to 0.47 in 2017. Yet the years 2015 and 2016 have shown an increasing trend. The mean doubling time for publications for the periods of 2013-2017 was 1.78 which has shown a steady growth of publications in DESIDOC Journal.

**Analysis of number of Pages of Articles**

**Table7: Year Wise Output of Number of Pages of Articles**

| S.NO | YEAR/VOL | 1-5 | 6-10 | ABOVE 10 | TOTAL | % |
|------|----------|-----|------|----------|-------|---|
| 1 | 2013/33 | 4 | 26 | 3 | 33 | 14.5 |
|   |          | 10.81 | 14.86 | 20.00 | | |
| 2 | 2014/34 | 13 | 45 | 5 | 63 | 27.7 |
|   |          | 35.14 | 25.71 | 33.33 | | |
| 3 | 2015/35 | 6 | 34 | 4 | 44 | 19.4 |
|   |          | 16.22 | 19.43 | 26.67 | | |
| 4 | 2016/36 | 7 | 27 | 2 | 36 | 15.9 |
|   |          | 18.92 | 15.43 | 13.33 | | |
| 5 | 2017/37 | 7 | 43 | 1 | 51 | 22.5 |
|   |          | 18.92 | 24.57 | 6.67 | | |
|   | TOTAL | 37 | 175 | 15 | 227 | 100 |

**Graph3: Analysis of number of Pages**

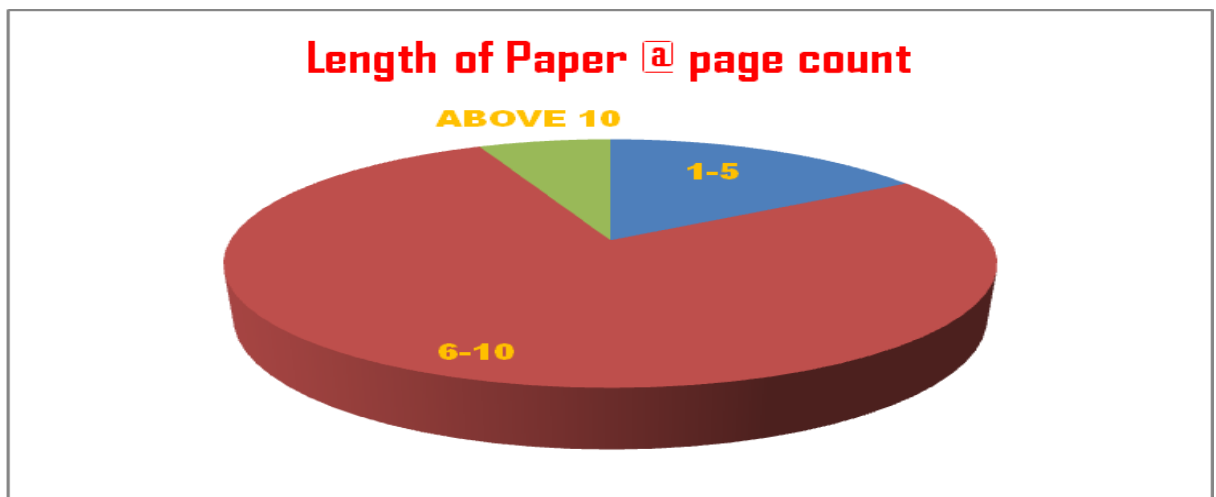

The table and figure indicates the Length of papers in terms of pages. The Maximum number of pages of papers published in DESIDOC journals during 2013 to 2017 were analysed and the result shows that the majority of articles covered their contents between 6 and 10 pages. It is witnessed by the data recorded in the years from 2013 to 2017 as a result the the highest score of 175 is observed in the range of 6 to 10 pages. It is followed by 1-5 pages which is accounted to 37 in total. More than 10 pages were accounted only 15 papers. This range is noted to be less in this study.

**Analysis of Subject distributions of the publication**

**Table8: Subject Distributions of the Articles Published**

| S.NO | MAJOR SUBJECTS | 2013 | 2014 | 2015 | 2016 | 2017 | TOTAL |
|---|---|---|---|---|---|---|---|
| 1 | Scientometrics, Bibliometrics | 11 | 18 | 10 | 1 | 11 | 51 |
| 2 | Webometrics | 1 | - | 2 | - | 2 | 5 |
| 3 | User survey | 3 | 6 | 4 | 9 | 9 | 31 |
| 4 | E-Resources | 3 | 9 | 2 | 5 | 8 | 27 |
| 5 | Information Seeking Behaviour | 2 | 2 | 1 | 1 | - | 6 |
| 6 | Knowledge Management | 2 | 2 | 3 | 3 | 1 | 11 |
| 7 | Library Services | 2 | 3 | 2 | 2 | 3 | 12 |
| 8 | ICT | 1 | 5 | 1 | 1 | 1 | 9 |
| 9 | Digital Libraries | 1 | 1 | 1 | 2 | 1 | 6 |
| 10 | Open Access | 2 | 1 | 2 | 1 | 3 | 9 |
| 11 | Library Automation | 1 | 2 | 1 | 2 | 2 | 8 |
| 12 | Search Engines | - | - | 2 | - | 1 | 2 |
| 13 | Social Networks | - | - | 1 | 2 | 1 | 3 |
| 14 | Others | 4 | 14 | 12 | 7 | 9 | 46 |
|  | TOTAL | 33 | 63 | 44 | 36 | 51 | 227 |

The articles covered in DESIDOC journals have been analysis on the basis of their subject coverage during the study period. The highest coverage of subject included in the journal is Scientometrics and Bibliometrics. The less concentrated subject of publication brought out by this journals is about Search Engine which is accounted to just 2. The major count of subject of Scientometrics and Bibliometrics was calculated to 51 articles. It is clear that the core concentration given by DESIDOC journals is Scientometrics and Bibliometrics.

## V. FINDINGS AND CONCLUSION

The data was collected by the online sources of DESIDOC website. The Archives of the journals were downloaded and the pdf of all the articles were analysed for the purpose of present study. The period of study taken for analysis is from 2013 to 2017. The findings revealed the following facts. The highest number of publications brought out by DESIDOC Journal of Library & Information Technology (DJLIT) was in the year 2014 and the total number articles at the highest were accounted to 63. The extent of research contributions by the authors is explained under authorship pattern and it was found that out of 227 articles, majority of articles were published by two authors and it was about 111 shows 50% percentages. It is also witnessed with the degree of collaboration at 0.69 percentages.

The page wise distribution of publications reveals the findings that most of the contributors have ranged their contents between 6 and 10 pages. It is about 175 in total. The total average number of authors per paper found in the study was 9.65 and the average productivity per author was 2.59. The Exponential Growth Rate was found to be 4.85 and also CAGR calculated was 9.1%, which shows a developmental sign of activity. The study period has witnessed a mean relative growth rate of 0.61. The mean doubling time for publications for the periods of 2013-2017 was 1.78. The subject core concentration of the journal observed in the study Scientometrics and that the present paper is an apt study supporting the field of study and this work regards the DESIDOC Journal of Library & Information Technology (DJLIT) for its effort in developing the study of science of science in the context of Information Science.